# Determination of positive anode sheath in anodic carbon arc for synthesis of nanomaterials

N. S. Chopra[1,2], Y. Raitses[1], S. Yatom[1], J. M. Muñoz Burgos[3]

[1] Princeton Plasma Physics Laboratory, Princeton University, Princeton, NJ, 08543, United States of America
[2] Department of Astrophysical Sciences, Princeton University, Princeton, NJ, 08544, United States of America
[3] Astro Fusion Spectre LLC, San Diego, CA, 92127, United States of America

**Abstract.** In the atmospheric pressure anodic carbon arc, ablation of the anode serves as a feedstock of carbon for production of nanomaterials. It is known that the ablation of the graphite anode in this arc can have two distinctive modes with low and high ablation rates. The transition between these modes is governed by the power deposition at the arc attachment to the anode and depends on the gap between the anode and the cathode electrodes. Probe measurements combined with optical emission spectroscopy (OES) are used to analyze the voltage drop between the arc electrodes. These measurements corroborated previous predictions of a positive anode sheath (i.e. electron attracting sheath) in this arc, which appears in both low and high ablation modes. Another key result is a relatively low electron temperature (~ 0.6 eV) obtained from OES using a collisional radiative model. This result partially explains a higher arc voltage (~ 20 V) required to sustain the arc current of 50-70 A than predicted by existing simulations of this discharge.

## 1. Introduction

Carbon nanomaterials are promising candidate materials for numerous chemical conversion and chemical storage applications. For instance, various forms of carbon nanotubes are desirable for storage of hydrogen for fuel cells or as catalysts for biorelated reactions [1–3]. The anodic carbon arc is a promising method for low-cost, high-volume synthesis of carbon nanomaterials, including graphene flakes, fullerenes, and single- and multi-walled carbon nanotubes [4–7]. Previous works have reported the formation of carbon nanomaterials in the arc periphery, on the chamber walls, and deposited on the cathode [8–11]. Such arcs are usually run in a background of atmospheric pressure helium gas. During the arc operation, carbon material is introduced into the arc plasma by the ablation of the graphite anode [5,7–9]. The anode ablation depends on the power balance at the anode, which among different factors, should be influenced by whether the anode sheath is electron-repelling (negative anode sheath) or electron-attracting (positive anode sheath) [7,12–16]. Another important feature of the carbon arc is the transition between so-called low and high ablation regimes which occurs at a certain arc current threshold. This current threshold can be affected by the anode diameter, gas pressure, and the gap between the anode and the cathode [7,15,17]. For low ablation, the arc current is below the current threshold The anode ablation rate changes insignificantly with respect to arc current in this mode. Above the current threshold, the ablation rate of the anode grows rapidly and nonlinearly with the arc current [7,14,17].

Recent theoretical studies have modeled the carbon arc plasma parameters and have predicted that the transition from low to high ablation mode could be explained by the presence of the background working gas (e.g. helium) impeding ablated carbon flow [15,16]. Models predict that this transition is also influenced by the sheath between the plasma and the anode (anode sheath) [15]. However, there is an unresolved discrepancy between these models and experiments. While experimentally determined discharge voltages are typically observed to be ~15-20 V, models of the arc underpredict the discharge voltage by ~50% [16]. It remains unclear whether this discrepancy in discharge voltage is due to a discrepancy in modeling of the anode sheath or to other energy loss mechanisms in the arc [16]. In this work, this discrepancy is addressed through careful measurements and detailed analysis of the plasma properties of the arc, including plasma potential and electron temperature using probes and optical emission spectroscopy, respectively.

Previous studies have determined floating and plasma potential in carbon arc discharges but did not account for ion collisions with background gas atoms [18,19]. Previous studies have modeled the effects of collisions on the interpretation of probe data. Several of these provide models relating the potential of a floating surface to the plasma potential [20–22]. The effect of collisions on the interpretation of current-voltage characteristics of a biased probe for determining the electron energy distribution function has also been investigated [23]. With knowledge of basic plasma parameters (electron density and electron temperature) and ordering of relevant length scales (mean free path of ions and electrons, electron Debye length, and probe radius), the plasma potential can be deduced from measurements of the floating potential of the probe. In this work, this approach is used to determine the plasma potential with respect to the anode and thereby, characterize the anode sheath. The electron temperature and plasma density are determined by optical emission spectroscopy (OES), in a similar fashion to [14], but with a more accurate collision radiative model. The effect of arc motion on the measured probe potential is considered by correlating the measured probe potential with fast-frame images. The effect of ion-neutral collisions on the probe floating potential is considered in the determination of the plasma potential.

Measured results indicate the existence of a positive anode sheath in both low and high ablation modes of the carbon arc.

The paper is organized as follows: the experimental setup of the arc, probe, and optical diagnostics is described in Sec. 2. The experimental procedure is described in Sec. 3. The measurements of the arc properties are described in Sec. 4. Sec. 5 discusses the determination of the anode sheath and implications of the measured anode sheath for anode ablation. In this section, an explanation is provided for part of the discrepancy between experimentally determined discharge voltage and the discharge voltage found in recent models of the carbon arc [16]. The conclusions are summarized in Sec. 6.

## 2. Experimental setup

The arc setup used for these experiments is shown in Figure 1 and described elsewhere [14]. The arc electrodes are placed vertically in the arc reactor chamber, which is equipped with a mechanical vacuum pump to evacuate the air before the experiment and maintain the buffer gas at sub-atmospheric pressure during the arc operation. The arc discharge is sustained with a Sorenson SGA100X100C-1AAA 100V/100A power supply, operated in a current regulated mode.

In the described experiments, the arc was operated with a background buffer gas of 500 torr He/$H_2$ gas mixture (95% He, 5% $H_2$ by concentration). The hydrogen was added to allow detection of the arc core via fast frame imaging of the hydrogen Balmer series $H_\alpha$ line. In addition, OES of the hydrogen Balmer series $H_\alpha$, $H_\beta$, $H_\gamma$, and $H_\delta$ lines was applied to determine the electron density, $n_e$, and the electron temperature, $T_e$. The anode and cathode electrodes are made from graphite and have diameters of 6.5 mm and 9.5 mm, respectively. As in previous studies reported in Refs. [7,14,17], the anode is placed on a positioning stage to enable the arc initiation and maintain a constant interelectrode gap during the arc operation.

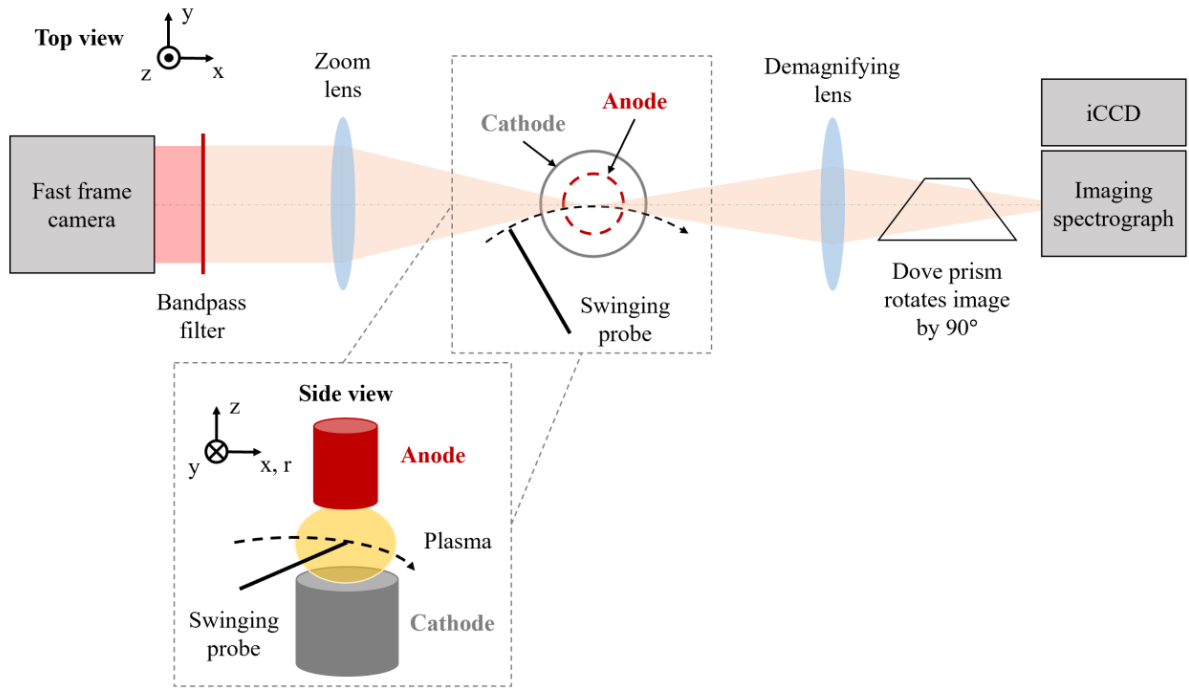

**Figure 1.** Diagram of the experimental setup.

An electrostatic probe diagnostic was implemented to determine the plasma potential, electron temperature, and plasma density in the arc. The probe was a 0.2 mm diameter W wire with 1.4 mm length of probe tip exposed to the plasma. The probe wire was housed in nested Alumina ceramic tubes, held together by Sauereisen brand 2 Aluseal adhesive cement paste. To prevent the probe tip from melting, the probe was mounted on a rotary feedthrough connected to an AX-18A servo motor. This swung the probe through the plasma in a plane perpendicular to the electrode axis (Figure 2), with a residence time in the plasma of ~50ms. A post arc inspection of the probe revealed no damage to the probe wire suggesting that its temperature did not exceed melting temperature of the tungsten (3695 K). At such temperatures, the thermionic electron emission takes place. However, even in the hot arc core with the expected plasma density of $10^{21} - 10^{22}$ m$^{-3}$, the maximum flux of thermionically emitted electrons from the wire at 3695 K should still be much smaller than the flux of the electrons from the plasma to the probe [14]. Under such conditions, the electron emission has a negligible effect on the floating potential of the probe [24].

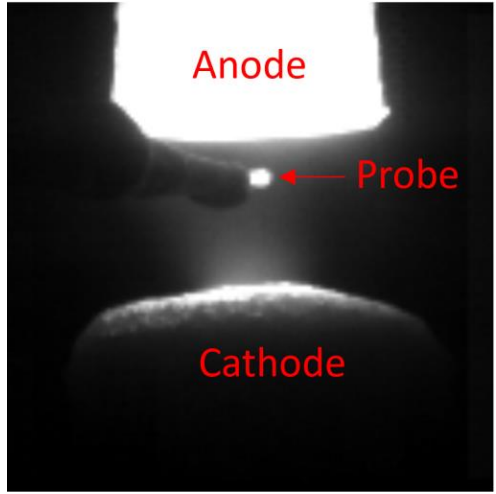

**Figure 2.** Fast frame image of the Langmuir probe in the core of the arc, with $H_\alpha$ filter employed to image arc core. Here the probe is imaged to be in the arc core, and the interelectrode gap is 4.5 mm.

Fast frame images of the arc are acquired with a Phantom v7.3 fast frame camera triggered on the same time base as the probe. The fast frame imaging optics consisted of a zoom lens and a $660 \pm 10$ nm bandpass filter, which enabled imaging of 656 nm $H_\alpha$ emission indicative of the position of the arc core.

OES was performed with a Horiba iHR-550 imaging spectrograph. A PI-MAX 3 Model 1024i iCCD camera was installed on the imaging spectrograph to collect the spectral images. The iCCD detector was set to a gate width of 500 ns, with 10000 accumulations per exposure, triggered at a rate of 1MHz leading to a total exposure time of 5 ms. The OES setup included a demagnifying lens and a dove prism to shrink and rotate the image of the arc by 90° respectively, enabling acquisition of radial profiles of the carbon arc. Acquiring radial OES profiles of the arc ensured the acquisition of the plasma parameters in the arc core.

## 3. Procedure

*Arc operation*

The arc was initiated by touching and then separating the biased electrodes. The arc current was determined by measuring the voltage drop across a 0.2 mΩ shunt resistor connected in series with the arc. In the described experiments, the arc current was varied from 50 to 65 A. This current range covered both low and high ablation modes. The transition between low and high ablation mode was roughly at 55 A. The continuous ablation of the anode and deposition onto the cathode during arc operation required active control of the electrode positioning to maintain a constant interelectrode gap. The anode was stationary, while the cathode axial position was continuously adjusted by a Velmex motor controller. The interelectrode gap distance was monitored during discharge with the fast frame camera and confirmed post discharge using calipers.

*Determination of arc V-I, anode ablation rate, and cathode deposition rate*

The discharge voltage $V_d$, defined as the difference in potential between the anode and cathode surfaces, was determined in a manner similar to Refs. [7] and [14]. An oscilloscope trace measures the voltage of the anode body several inches from the surface of the anode $V_a^*$, referenced to the grounded chamber; because of anode ablation, it is difficult to measure the potential of the anode surface directly during arc operation. Moreover, as the current flows through the electrodes, a voltage drop establishes over the finite resistivity electrodes themselves. For a given discharge current, the voltage drop across the anode and cathode electrodes is determined pre- or post-discharge by shorting the electrodes and measuring the resulting voltage drop; this shorted voltage will be denoted as $V_{short}$. The residual voltage drop from the cathode to ground, $V_c$, is determined by attaching a wire to the cathode. The discharge voltage is then determined by subtracting the shorted electrode voltage from the measured anode body potential and adding the cathode voltage drop to ground, $V_d = V_a^* - V_{short} + V_c$.

The average anode ablation rate and average cathode deposition rate were determined in a similar fashion to Refs. [7,14], by weighing the electrodes pre- and post-discharge. The electrode was first weighed before running a discharge. Then the electrode was installed in the chamber and the arc was run for a given interelectrode gap and discharge current for 60 seconds. Finally, the electrode was weighed again. The difference in mass of the electrode pre- and post- discharge divided by the duration of the arc run gives the time averaged rate of mass change of the electrode. To obtain statistically significant results, this procedure was repeated 3 times for each interelectrode gap and discharge current.

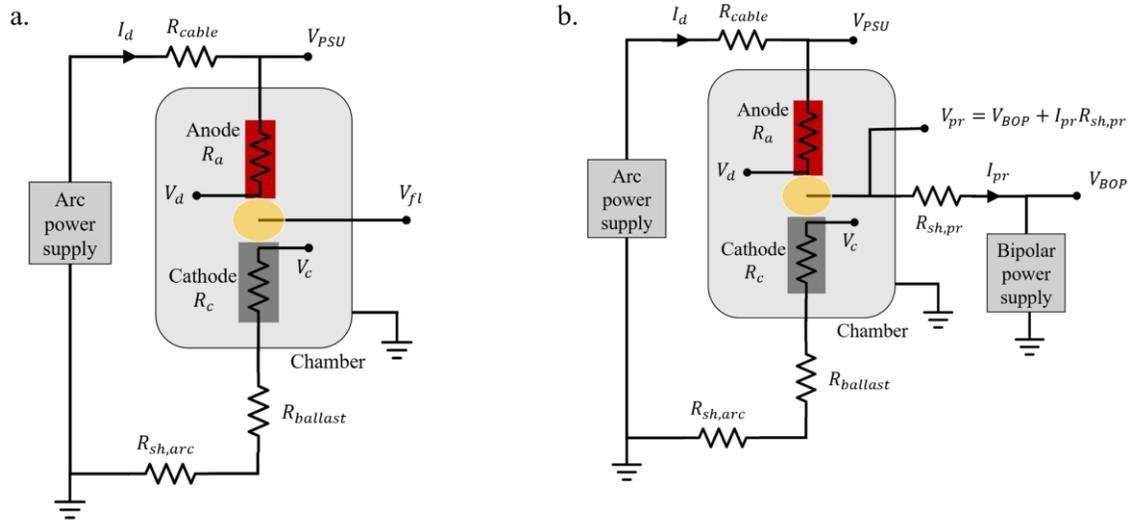

**Figure 3.** Electrical circuitry schematic of carbon arc experiment, with floating probe (a) and swept Langmuir probe (b) circuitry included.

*Probe diagnostic*

During experimental runs with the probe, the interelectrode gap was chosen to be 4.5±0.3 mm, a significant increase from the interelectrode gap used and modeled in previous works in which interelectrode gaps were kept to the 1-3 mm range [14,16]. A gap of 4.5 mm was the minimal interelectrode spacing at which the probe motion was reproducible without the probe tip colliding with the arc electrodes. This gap also reduced the probe-induced disturbance of the arc operation. The probe induced disturbance was characterized by measuring and confirming that the arc current did not deviate by more than 5% from its nominal value upon probe insertion into the plasma.

The probe was configured in two ways. The first configuration was a floating probe, where the potential of the floating probe $V_{pr}$ was determined using a Teledyne Lecroy AP031 differential probe referenced to the grounded chamber. To test reproducibility, $V_{pr}$ was measured in the arc core at midplane for 3 independent trials at each discharge current measured. The core plasma potential at the probe $V_{pl,pr}$ was then determined from the acquired floating potential at the arc core, also using the electron temperature determined via OES of the arc. As in Figure 1, let $r$ denote the radial coordinate relative to the center of the arc, with $r = 0$ mm as the center of the carbon arc plasma core. The floating potential of the probe at $r = 0$ was determined by correlating the probe signal to instants in time when the probe tip was imaged to be in the center of the arc core, determined by fast frame imaging of the arc through an $H_\alpha$ filter. The core position was taken to be the radial position of maximum intensity in the radial direction. The anode sheath voltage $V_{a,sh}$ is determined as

$$V_{a,sh} = V_a - V_{pl,a} \tag{1}$$

where $V_a$ is the potential of the anode surface and $V_{pl,a}$ is the plasma potential next to the anode. The plasma potential next to the anode is related to the plasma potential at the probe measurement location $V_{pl,pr}$ via

$$V_{pl,a} = V_{pl,pr} + V_{col} \tag{2}$$

where $V_{col}$ is the voltage drop over the plasma column to the probe measurement location (Appendix C). The plasma potential at the probe is in turn related to the measured probe potential $V_{pr}$ by

$$V_{pl,pr} = V_{pr} + V_{fl} \tag{3}$$

where $V_{fl}$ is the voltage drop over the floating probe sheath and presheath

$$V_{fl} = V_{sh} + V_{pre} \tag{4}$$

For typical plasma parameters in the hot core of the carbon ($0.5 - 1.0$ eV, $10^{21} - 10^{22}$ m$^{-3}$), the Debye length is $\lambda_{D_e} \sim 0.1$ μm, which is much smaller than the ion-neutral mean-free-path $\lambda_{in} \sim 1$ μm. Therefore, the sheath between the plasma and the probe can be assumed to be collisionless and has a sheath voltage drop of

$$V_{sh} = \frac{1}{2}\frac{kT_e}{e}\ln\left(\frac{M_i}{2\pi m_e}\right) \tag{5}$$

where $m_e$ and $M_i$ are electron and ion masses and $e$ is the charge of the electron [25]. For carbon ions, Eq. (5) gives $V_{sh} = 4.08 T_e$. For a collisionless presheath where ion energy is conserved, the presheath potential drop is $V_{pre} = \frac{1}{2}\frac{kT_e}{e}$. However, for a collisional presheath, where ions can exchange momentum with background neutrals, Ref. [22] showed that the presheath drop can be much larger, $V_{pre} = \alpha\frac{kT_e}{e}$, where $\alpha \geq 0.5$ is a factor that accounts for the extra voltage needed to compensate for the momentum transfer between ions and neutrals as ions are accelerated by the presheath to the Bohm velocity. In this work, $\alpha \sim 4$ (Appendix A.I). Finally, an expression for the anode sheath as a function of experimentally determined parameters $V_a, V_{col}, V_{pr}, \lambda_{in}$, and $T_e$ is

$$V_{a,sh} = V_a - V_{col} - V_{pr} - V_{sh}(T_e) - V_{pre}(\lambda_{in}, T_e) \tag{6}$$

The second probe configuration was a sweeping Langmuir probe, where the probe was biased with a 10 kHz sinewave output by a Kepco BOP 36-6M bipolar operational power supply (BOP). The input waveform that was amplified by the BOP was generated by a Rigol DG 2041A waveform generator. The probe collected IV traces at different radial positions in the plasma relative to the arc core center position. The current drawn by the probe was determined by measuring the voltage across a shunt resistor $R_{sh,pr}$ placed between the probe and the BOP. The resulting probe voltage was $V_{pr} = V_{BOP} + I_{pr}R_{sh,pr}$. The shunt resistor was chosen to be small relative to the anticipated sheath impedance [26]. A 2 Ω shunt was used when measuring ion saturation current, and a 18.2 Ω shunt was used when measuring electron saturation current.

The ion saturation current collected from probe measurements of the arc are used to determine a radial profile of the plasma density in the arc. The plasma density was deduced from the measured ion current using a model of the ion saturation current to the probe in collisional plasmas such as the carbon arc plasma [27]. In this model, ions diffuse through the background neutral carbon atoms in the arc to the probe surface, and all species (electrons, ions, and neutrals) are locally thermally equilibrated to a temperature $T$ with a total pressure $p$. In addition, the contribution of ionization in the probe presheath to the ion saturation current is considered. For carbon ions diffusing through neutral carbon atoms, the ion saturation current $I$ to the probe is given as

$$I = A\frac{eDn_e}{L_{iz}}\left(0.71 + 1.3\left(\frac{L_{iz}}{r_p}\right) + 0.44\left(\frac{n_e}{n_a}\right)^{0.77}\right) \tag{7}$$

where $A$ is the probe surface area, $D = 2T/[(M/2)S_{ia}v_{th}(p/T)]$ is the ambipolar ion diffusion coefficient, $M$ is the mass of a carbon atom, $S_{ia}$ is the carbon ion-atom collision cross section, $v_{th}$ is the thermal velocity of carbon ions and atoms, $L_{iz} = \sqrt{D/(\beta n_e^2)}$ is the characteristic ionization length for carbon, $\beta$ is the volumetric recombination coefficient for carbon, $r_p$ is the probe radius, $n_e$ is the bulk electron density, and $n_a$ is the bulk carbon neutral density. The value used for $\beta$ here was taken from Ref. [28]. For a measured ion saturation current, Eq. (7) is then solved numerically for $n_e$.

*OES methods*

Balmer series spectral line intensities for the $n = 3 \rightarrow 2$ $(H_\alpha)$, $4 \rightarrow 2$ $(H_\beta)$, $5 \rightarrow 2$ $(H_\gamma)$, and $6 \rightarrow 2$ $(H_\delta)$ principle quantum number transitions of the Hydrogen atom were collected. The Balmer series data was used to determine the electron temperature in two different ways. The first method assumed the hydrogen states for $n \geq 2$ were in pLTE, as supported by Griem's criterion [29], and the standard Boltzmann diagram method was used to determine $T_e$. The second method relaxed the pLTE assumption and allowed radiation, electron impact ionization, and three-body recombination to play a role in populating the excited states of hydrogen. In this method a collisional radiative model (CRM) was used to determine the relative emissivities of the $H_\alpha$, $H_\beta$, and $H_\gamma$ lines. The determined $T_e$ of the arc core converged within error between the pLTE and CRM methods, supporting the hypothesis that the arc plasma is at a high enough density to assume pLTE for the Balmer series. Finally, the core plasma density $n_e$ was determined via OES in a method similar to [30], by performing a Voigt profile fit to the $H_\alpha$ line and extracting the Stark broadening contribution to the profile FWHM. These methods are detailed in Appendix B.

Plasma opacity effects to each of the Balmer series lines were estimated to be negligible for the range of parameters considered in the carbon arc [14,30]. Time resolved measurements, taken with total exposure time of only 0.1 ms, were also obtained, but only for $H_\alpha$ and $H_\beta$ lines due to fundamental detector limits. The time averaged data that excluded the $H_\gamma$ and $H_\delta$ lines was compared to time resolved data and found similar results $T_e$ that agreed within $0.05\ eV$. Therefore, it was concluded that the effects of time averaging on the prediction of $T_e$ were negligible.

## 4. Results

*Arc V-I and ablation rate*

Ablation and deposition rate data was collected for 2.0 mm and 4.5 mm interelectrode gaps, shown in Figure 4. The ablation and deposition rates for the 2.0 mm gap case were found to agree well with previous work [14]. The anode ablation rate and cathode deposition rates showed dependence on gap distance. A possible explanation of the diminished anode ablation rate in the larger 4.5 mm gap case is provided by [15]; as the gap increases, less of the thermal radiation emitted from the cathode surface reaches the anode surface, resulting in a lower anode surface temperature and hence lower anode ablation rate. The lower cathode deposition rates in the 4.5 mm gap case are attributed to increased radial losses of ablated material. As the interelectrode gap increases, more of the ablated anode material can diffuse radially outward and escape depositing on the cathode surface. Still, there was a clear increase in ablation rate between low discharge current (50A) and high discharge current (65A) regimes, consistent with a transition from low to high ablation mode.

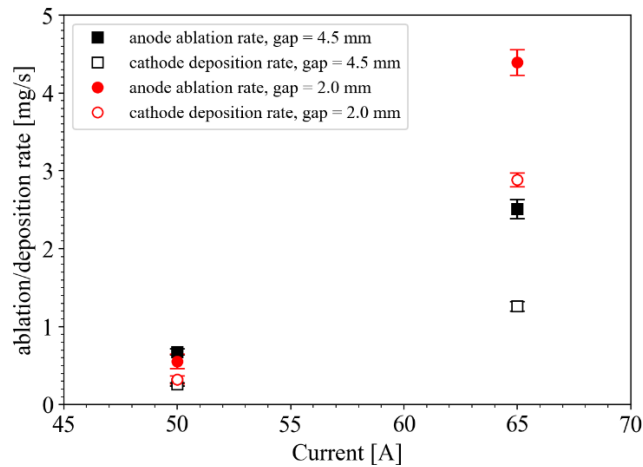

**Figure 4.** Anode ablation and cathode deposition rates for 4.5mm gap carbon arc.

The arc V-I was determined for 1.0, 2.0, 3.0, and 4.5 mm gaps, shown in Figure 5. The arc discharge voltage was found to be ~22V for a 4.5 mm interelectrode gap. A notable feature was that the arc V-I showed a strong dependence on gap distance in the low ablation regime, but lost this dependence in the high ablation regime. This may be explained by two competing effects. For a fixed discharge current, elongation of the interelectrode gap increases the length of the plasma column, which increases the voltage drop over the plasma column due to a larger resistance of the plasma column. However, as confirmed by the OES and probe measurements described in the following sections, in the high ablation mode the plasma density is higher, meaning the column resistivity decreases. This has the effect of reducing the contribution of column elongation to the discharge voltage.

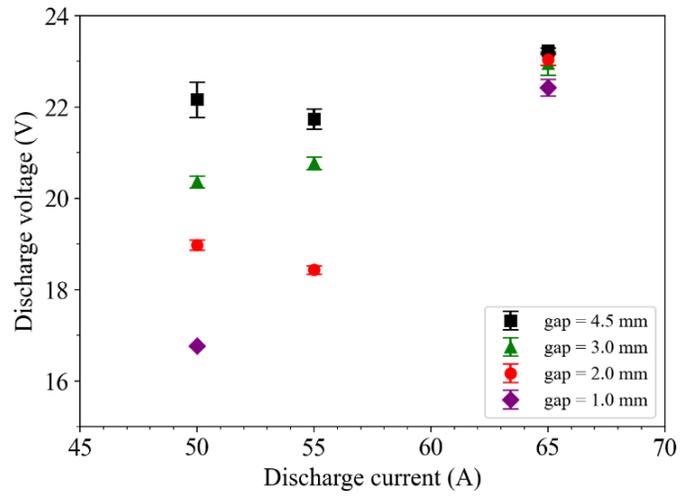

**Figure 5.** Discharge voltage versus discharge current.

*OES measurements*

Measurements of the hydrogen Balmer series yields arc core electron temperature (Table 1, Figure 6) and core plasma density (Table 2, Figure 7). Values of $T_e$ determined by OES agree within error with the $T_e$ determined by Langmuir probe data. A typical acquisition of the $H_\alpha$ line can be found in Appendix B.II. Importantly, when running the arc in the 2.0 mm gap, 55 A case, the electron temperature was found to be $T_e = 0.57 \pm 0.06$ eV, roughly 0.2 eV smaller than $T_e$ reported in Ref. [14] for an identical gap and discharge current. This has important implications in modeling of the arc, and is addressed further in the Discussion section of the paper.

| $T_e$ [eV] | $I_d = 50$ A | $I_d = 55$ A | $I_d = 65$ A |
|---|---|---|---|
| gap = 2.0 mm | | $0.57 \pm 0.06$ | |
| gap = 4.5 mm | $0.58 \pm 0.06$ | | $0.65 \pm 0.06$ |

**Table 1.** Results of core electron temperature as measured by OES, in units of eV.

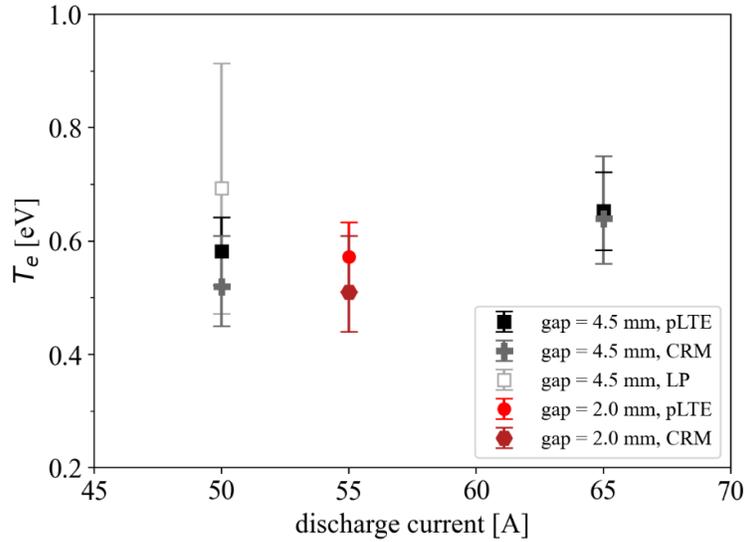

**Figure 6.** Core electron temperature determined by Boltzmann diagram method assuming pLTE and CRM method of time averaged OES Balmer series lines in the midplane of the carbon arc. Also plotted is core electron temperature as determined by Langmuir probe.

Saha ionization equilibrium is described by the following equation

$$n_e^2 = (n_a - n_e) \frac{2}{\lambda_{dBr}^3} \frac{g_1}{g_0} \exp(-E_i/kT_e) \tag{8}$$

where $n_a$ is the density of neutral carbon gas atoms, $g_0 = 9 \times 2 = 18$ is the degeneracy of the carbon ground state, $g_1 = 6 \times 2 = 12$ is the degeneracy of the ground state of a singly ionized carbon atom, $\lambda_{dBr} = (h^2/2\pi m_e k_B T_e)^{0.5}$ is the thermal de Broglie wavelength, and $E_i$ is the first ionization energy of the carbon atom. Experimentally determined plasma density was found to agree within error with values of $n_e$ calculated by Eq. (8) (Figure 8). This provides supporting evidence that the arc is in Saha equilibrium.

| $n_e$ [m$^{-3}$] | $I_d = 50$ A | $I_d = 55$ A | $I_d = 65$ A |
|---|---|---|---|
| gap = 2.0 mm | | $(7 \pm 1) \times 10^{21}$ | |
| gap = 4.5 mm | $(5 \pm 1) \times 10^{21}$ | | $(1.5 \pm 0.3) \times 10^{22}$ |

**Table 2.** Results of core plasma density measured by OES, in units of $m^{-3}$.

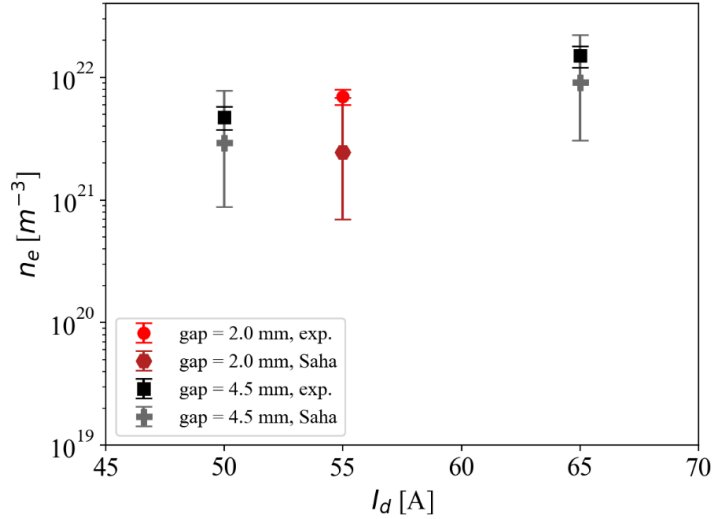

**Figure 7.** Experimentally determined electron density at different gaps (labeled "exp."), compared to calculation of plasma density determined by Saha equation (labeled "Saha").

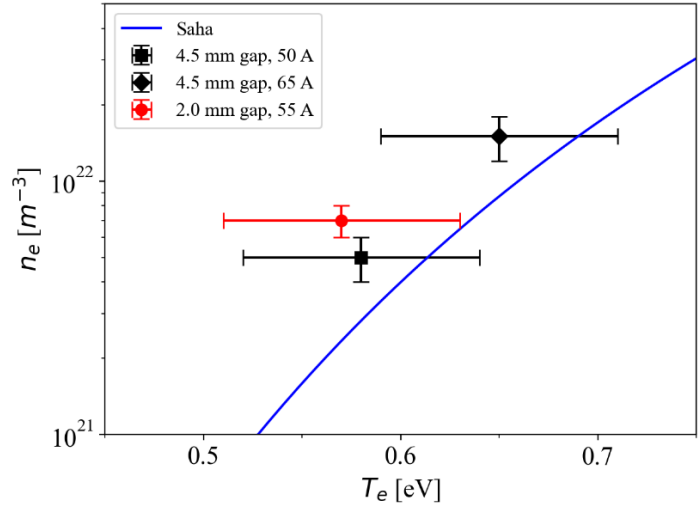

**Figure 8.** Plasma density as predicted by the Saha equation (Eq. (8)). Also plotted is experimentally determined electron temperature (using pLTE assumption) and plasma density.

*Probe measurements*

The floating potential of the probe in the arc core at midplane did not vary significantly between low and high ablation cases and was found to be ~13 V (Figure 9).

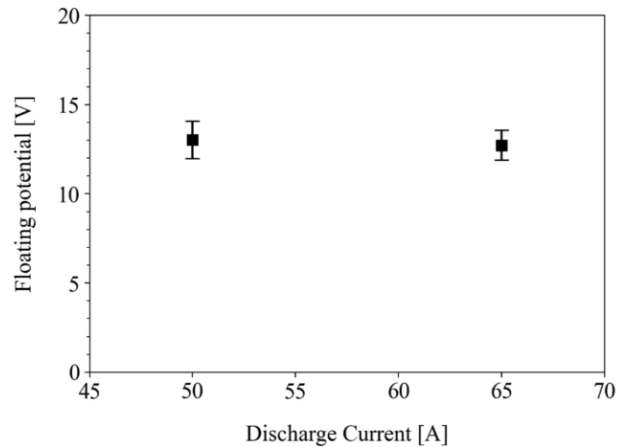

**Figure 9**. Floating potential of the probe when located in the arc core, measured at midplane, in low and high ablation regimes, gap = 4.5 mm.

Ion saturation current at several radial locations was determined as in Ref. [31] by determining a linear fit to the ion saturation regime of the Langmuir probe I-V trace and extrapolating this fit to the plasma potential. The ion saturation region was determined as the region of the probe I-V to the left of the I-V inflection point (Figure 10). This inflection point is indicative of the transition from the probe collecting only ion saturation current to collecting Boltzmann electrons in addition to ion saturation current. Electron temperature at $r = 0$ mm was also calculated from the negatively biased probe I-V (Figure 6). Here, it was assumed that in the Boltzmann electron transition region of the probe I-V the electron current increases exponentially with the probe bias voltage, $(I_{probe}(V_B) - I_{i,sat}) \propto \exp\left(\frac{eV_B}{kT_e}\right)$. Finally, the plasma potential was also calculated from analyzing a positively biased probe that collected mainly electron current. Plasma potential was calculated as the probe bias voltage for which the positively biased probe I-V had an inflection point (Figure 11). This is not to be confused with the determination of plasma potential using the floating probe and electron temperature measurements, results of which are described in the following section.

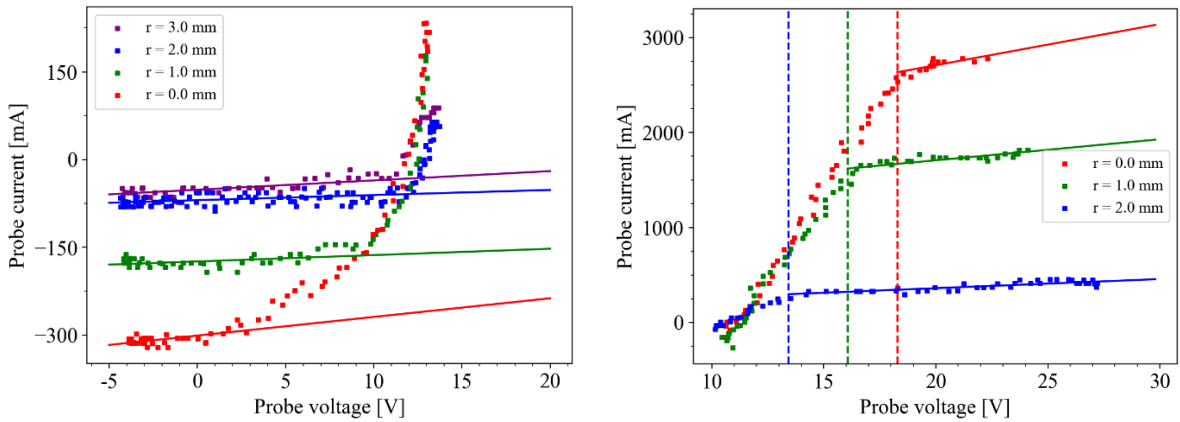

**Figure 10.** Typical ion saturation (left) and electron saturation (right) traces collected at various locations relative to the arc core. Also plotted are linear fits to the saturation regions (solid lines) and location of IV 'knee' (dashed lines). (50 A)

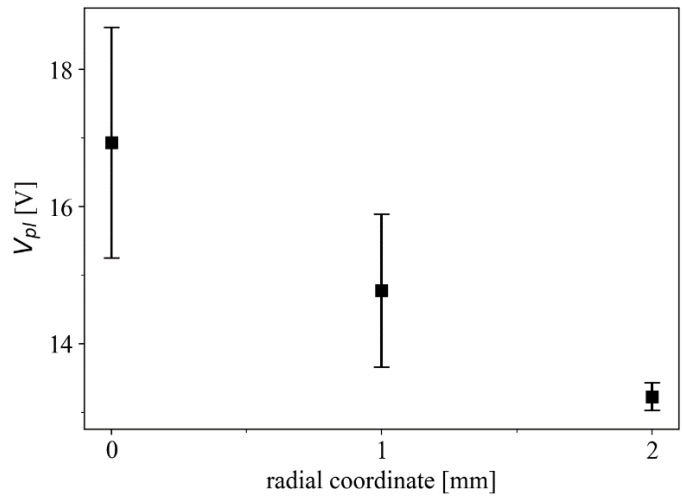

**Figure 11.** Plasma potential as determined by the swept Langmuir probe data, measured at midplane. (50 A, 4.5 mm gap)

The model of Ref. [27] applied to the measured ion saturation current data at the arc core shows good agreement with $n_e$ as determined by OES (Figure 12). This suggests that the effects modelled by Ref. [27] are important considerations for future experimentation using negatively biased probes in the carbon arc. These are, namely, that the diffusion of ions against background neutrals and the production of ions in the probe presheath can significantly contribute to the ion saturation current, affecting the prediction of the bulk plasma density from ion saturation current.

Plasma density at the arc core as determined by probe and OES methods in this work were found to be $n_e = (5 \pm 1) \times 10^{21}$ m$^{-3}$, which was slightly lower than reported in previously experimentally determined core plasma density of $n_e = 8 \times 10^{21}$ m$^{-3}$ in the 2 mm interelectrode gap carbon arc in Ref. [14]. This slight discrepancy is attributed to the larger interelectrode gap in this work leading to larger radial plasma losses from the arc core. Both this work and Ref. [14] reported significantly lower core plasma density than modeled in Ref. [16], which simulated the core plasma density to be $n_e = 2 \times 10^{22}$ m$^{-3}$.

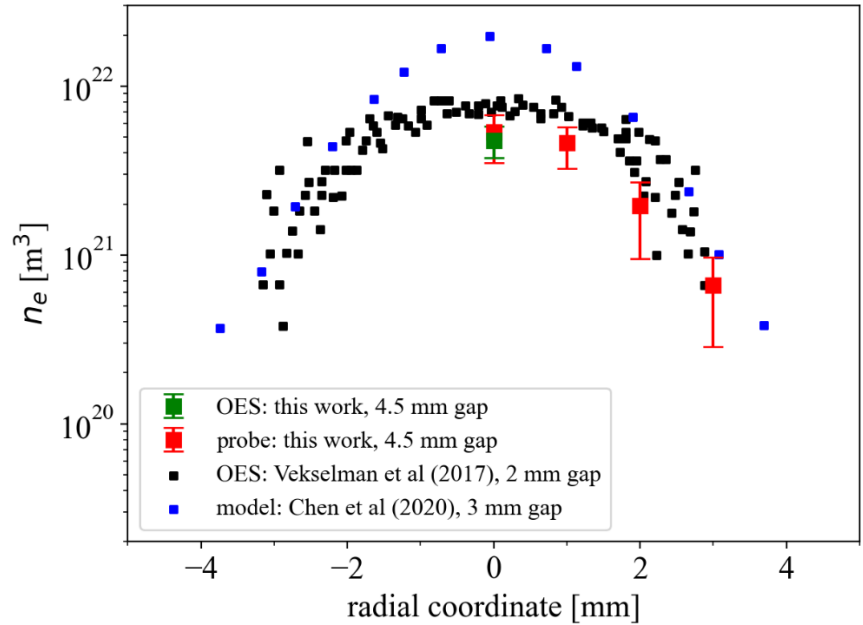

**Figure 12.** Radial plasma density profile for a 50A arc as determined by probe and spectroscopic methods in this work and in Ref. *[14]*. Also plotted is the plasma density profile predicted by Ref. *[16]*.

*Determination of plasma potential and anode sheath potential*

Using the experimentally determined values for $T_e$ and probe floating potential and assuming a collisional probe presheath, the plasma potential at the arc core at midplane is found to be $V_{pl} = 17.9 \pm 1.2$ V at $I_d = 50$ A and is $V_{pl} = 18.1 \pm 1.0$ V at $I_d = 65$ A. Thus, the measured plasma potential in the arc core at midplane does not vary much between low and high ablation mode (Figure 13).

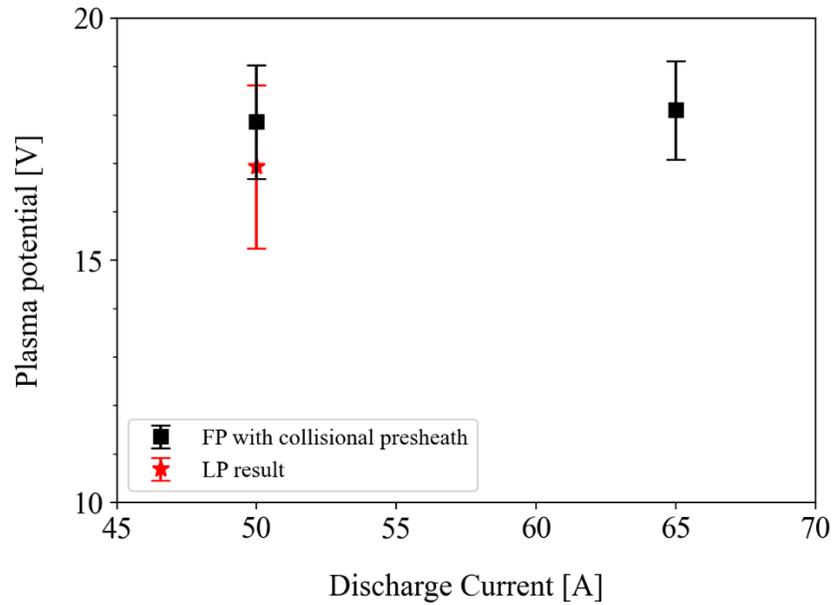

**Figure 13.** Plasma potential in the arc core, measured at midplane, as determined by the floating probe (FP) measurement with a collisional presheath and the knee of the Langmuir probe (LP) IV characteristic.

Using the measured anode potential and estimated plasma column drop, the anode sheath is calculated using Eqn. (6) to be $V_{a,sh} = 3.3 \pm 1.6$ V at $I_d = 50$ A and $V_{a,sh} = 7.2 \pm 1.6$ V at $I_d = 65$ A for the collisional probe presheath model. Note that despite the measured plasma potential being constant within error between the low and high ablation cases, the anode sheath potential significantly increases from low to high ablation mode. This happens because the arc voltage increases from low to high ablation modes. The anode sheath drop is measured to be positive in both low and high ablation regimes (Figure 14).

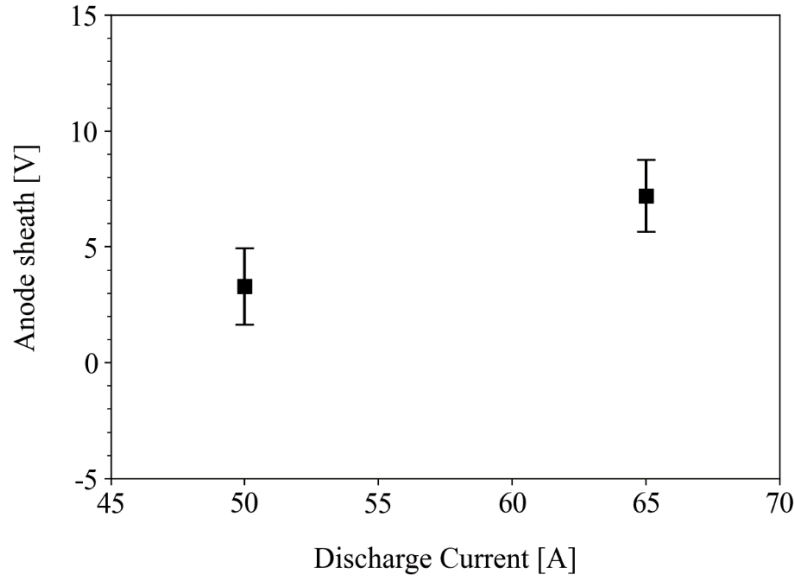

**Figure 14.** The voltage drop across the anode sheath-presheath as determined by collisional presheath models. (4.5 mm gap)

## 5. Discussion

*Positive anode sheath and implications for anode ablation*

As shown in Figure 14, the anode sheath is positive in both low and high ablation modes and significantly increases by several volts from a low to high ablation regime. These values of the anode sheath are several volts higher than used in recent modeling studies of a similar carbon arc [15,16]. The positive anode sheath can play an important role in delivering heat to the anode surface by accelerating electrons towards the anode surface [7,13,15].

The anode sheath and ablation rate data determined in this work were compared to a recently developed model of the anode ablation rate [15]. In the model, the anode ablation rate is determined by the power balance at the anode surface (Eq. (9)). This power balance assumes the cooling of the anode surface is due to ablation, heat conduction through the anode body, and blackbody radiation from the anode surface, and that these processes are balanced by the heating of the anode surface due to electrons gaining energy over the anode sheath next to the anode surface and work function voltage drop at the anode surface, thermal flux of electrons, and cathode radiation impinging on the anode surface. This is expressed as [15]

$$\pi r_a^2 g_{abl}(T_a)L + C_1 T_a^{2.5} r_a^{1.5} + C_2 r_a^2 T_a^4 = V_{eff}I + C_3 r_a^2 \tag{9}$$

$$V_{eff} = \max(V_{sh,a}, 0) + V_w + 2.5\frac{kT_{e,a}}{e}$$

$$C_1 = \pi\sqrt{\frac{4}{5}\sigma\varepsilon_a\lambda_a}$$

$$C_2 = \pi\sigma\varepsilon_a\alpha_r$$

$$C_3 = \pi\sigma\varepsilon_a\varepsilon_c F_{c\to a} T_c^4$$

$$\alpha_r = F_{a\to c}(1-\varepsilon_c)F_{c\to a}\varepsilon_a$$

Here $r_a$ and $r_c$ are the anode and cathode radii, $g_{abl}$ is the anode ablation rate per unit area of anode surface, $T_a$ and $T_c$ are the anode and cathode surface temperatures, $T_{e,a}$ is the electron temperature at the anode surface, $\varepsilon_a$ and $\varepsilon_c$ are the emissivities of the anode and cathode surfaces, $\lambda_a$ is the thermal conductivity of the anode, and $F_{a\to c}$ is a geometrical view factor between the cathode and anode front surfaces. The effective voltage $V_{eff}$ contributing to anode heating consists of the work function of graphite $V_w$, the thermal energy of electrons impinging on the anode $2.5\,kT_{e,a}/e$, and the anode sheath voltage $V_{a,sh}$. Here, $T_{e,a}$ is the bulk electron temperature $T_e$ as determined from OES. Note that the total ablation flux is given as $G_{abl} = \pi r_a^2 g_{abl}$. To solve Eq. (9) the anode surface temperature $T_a$ can be determined by assuming the ablated carbon pressure is determined by the Clausius-Clapeyron relation and is limited by the background helium gas. [15]:

$$\frac{1}{T_a} = \frac{1}{T_{sat}} - \frac{k}{Lm_C}\ln\left(\frac{g_{abl}}{p}\sqrt{\frac{2\pi kT_{sat}}{m_C}} + \left[1-\exp\left(-\frac{g_{abl}}{g_0}\right)\right]\right) \tag{10}$$

Solving Eqs. (9) and (10), an expression for the total ablation flux as a function of the anode sheath $G_{abl}(V_{a,sh})$ can be obtained. The theoretical dependence of $G_{abl}$ on $V_{sh,a}$ for a 65A arc is plotted in Figure 15. Here, the experimental ablation rate and anode sheath determined in this work are also shown. Apparently, the theory of Ref. [15] with the input of the anode sheath from these experiments overpredicts the ablation rate for the experimentally determined $V_{sh,a}$.

Suppose, instead, that only a fraction $\gamma$ of the power dissipated by the anode sheath serves to heat the anode surface, such that a "power loss" $(1-\gamma)IV_{sh,a}$ is spent on other external processes. This power loss could be characterized as power invested in heating, dissociating, or ionizing already ablated carbon atoms and molecules or plasma losses on particles spalled from the anode [10,13,32]. Because of this power loss, the anode receives only $\gamma IV_{sh,a}$ of ohmic power, and $V_{eff}$ is modified to be

$$V_{eff} = V_w + 2.5\frac{kT_{e,a}}{e} + \max(\gamma V_{sh,a}, 0) \tag{11}$$

As a result, for a given ablation rate, a larger $V_{sh,a}$ is needed to provide adequate heat to the anode surface. The ablation rate $G_{abl}(V_{sh,a})$ for $\gamma = 0.75$ is also shown in Figure 15, and is shown to agree well with the experimentally determined $G_{abl}(V_{sh,a})$. This suggests that only a fraction of the electron energy gained in the positive anode sheath is expended on heating the anode surface in the high ablation mode.

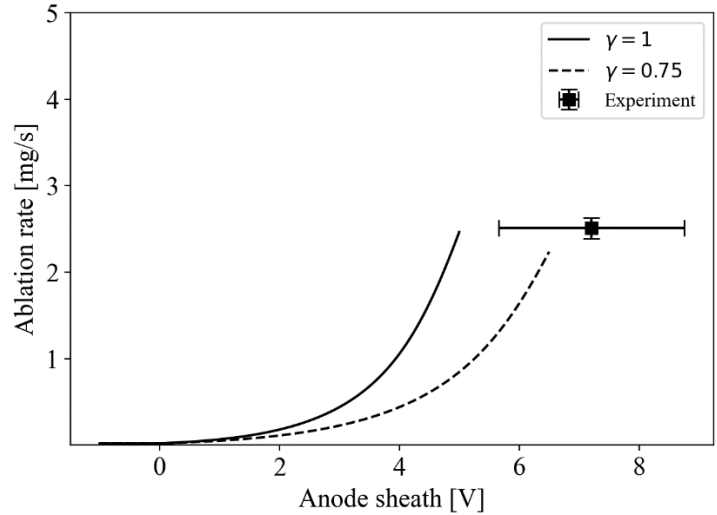

**Figure 15.** Theoretical and experimental ablation rate versus anode sheath potential in a high ablation ($I_d = 65A$) carbon arc, for 4.5 mm gap. Shown are the predicted ablation rate dependence on anode sheath assuming $\gamma = 1$ (solid line) and $\gamma = 0.75$ (dashed line). Also plotted is experimentally determined ablation rate and anode sheath data for 65A (square data point).

*Power dissipation in the anode and cathode sheaths*

The arc current is an easily and conveniently measured parameter, but the anode ablation is governed by the power deposited to the anode surface. As evident in Eq. (9), the anode ablation is a function of both the arc current and the anode sheath. As a result, the threshold for the transition from low to high ablation mode is more appropriately characterized by a transition over a threshold value of power deposition at the anode (Figure 16).

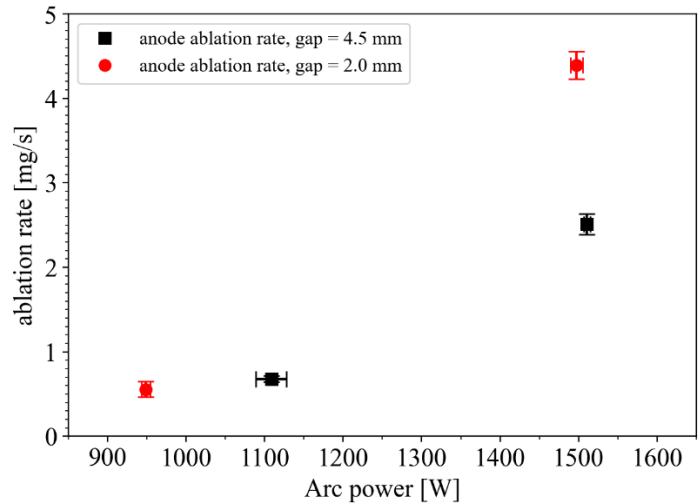

**Figure 16.** Anode ablation rate vs. arc power.

The total ohmic power dissipated in the arc $P_d = I_d V_d$ is comprised of the power dissipated in the anode sheath $P_a$, plasma column $P_{col}$, and cathode sheath $P_c$. This relation is expressed as

$$P_d = P_a + P_{col} + P_c = I_d V_a + I_d V_{col} + I_d V_c$$

where $I_d$ is the discharge current, $V_d$ is the discharge voltage, $V_a$ is the anode sheath, $V_{col}$ is the column voltage drop, and $V_c$ is the cathode sheath. The quantities $V_d$, $V_{col}$, and $V_a$ are all experimentally determined in this work in both low and high ablation regimes, at 50 A and 65 A. Therefore, the change in cathode sheath power deposition as the arc transitions from low to high ablation mode, $\Delta P_c$, can be calculated from the equation

$$\Delta P_d = \Delta P_a + \Delta P_{col} + \Delta P_c$$

Referring to
Table 3, the power deposition in the anode and cathode sheaths both increase and are of similar order. This is accompanied by the increased conductivity of the arc column in high ablation mode, indicated by the negative change in power deposition along the arc column.

| $\Delta P_d$ (W) | $\Delta P_a$ (W) | $\Delta P_{col}$ (W) | $\Delta P_c$ (W) |
|---|---|---|---|
| $402 \pm 20$ | $304 \pm 134$ | $-215 \pm 88$ | $314 \pm 162$ |

**Table 3.** Change in power deposition throughout the arc between 50 A and 65 A discharge current cases. Units are in Watts (4.5 mm gap).

*Discharge voltage discrepancy between model and experiment*

Recent models of the carbon arc show a discrepancy with experiment in determining arc voltage, underpredicting the discharge voltage by ~10V [14,16]. This is in spite of a good agreement between experiments and models regarding $T_e$ which was found to be $T_e = 0.8 \pm 0.1$ eV for the low ablation regime [14,16]. However, the results of this paper find a significantly lower electron temperature, $T_e = 0.58 \pm 0.06$ eV. Both this work and Ref. [14] used OES line ratio methods to determine electron temperature. The analysis in Ref. [14] also included a collisional radiative correction factor in the expression for line ratios used in Boltzmann diagram method. The use of this CRM factor is not necessary and, in fact, can overpredict $T_e$. This implies that the electron temperature used in recent models such as in [16] are larger than the actual experimentally observed value for $T_e$ reported in this work. The lower $T_e$ may explain part of the discharge voltage discrepancy previously seen between model and experiment, i.e. between Refs. [16] and [14]. This discrepancy may occur because the plasma conductivity reduces as $T_e$ and $n_e$ reduce.

The expression for plasma conductivity in the arc column $\sigma$ used in Ref. [16] is given by

$$\sigma = \frac{n_e e^2}{m_e(\nu_{e,a}(T_e) + \nu_{e,i}(n_e, T_e))}$$

where $\nu_{k,j} = \frac{4}{3}\sqrt{\frac{8kT_{kj}}{\pi m_{kj}}} C_{kj} Q_{kj} n_j$, is the effective binary collision frequency between species $k$ and $j$, $T_{kj} = \frac{(m_k T_j + m_j T_k)}{m_k + m_j}$ is the corresponding binary temperature, $m_{kj} = \frac{m_k m_j}{m_k + m_j}$ is the binary mass, and $Q_{kj}$ is the binary collision cross section. The term $C_{kj}$ is the kinetic coefficient of binary collisions between species $k$ and $j$, which is typically of order unity. Detailed calculations of $C_{kj}$ are discussed in [33,34]. Here it is assumed that the heavy species are in equilibrium, and the temperature of the carbon neutrals and ions is taken to be $T_C = 6000$ K as modeled in Ref. [16] for a 1.5 mm interelectrode gap.

Considering cases from computer simulation and experiment at the same discharge current $I_d$, Ohm's law is written as

$$I_d = \frac{V_{col,mod}}{R_{col,mod}} = \frac{V_{col,exp}}{R_{col,exp}}$$

where $V_{col}$ is the voltage drop across the plasma column, $R_{col}$ is the plasma column resistance, and the subscripts $mod$ and $exp$ refer to parameters from the model used in Ref. [16] and the experiment of this work, respectively. Rearranging this equation into a ratio of resistances, one can eliminate the geometrical contributions to the plasma column resistance ($R = \frac{AL}{\sigma}$, where $A$ is the area through which current is conducted and $L$ is the length of the column. That is, assuming $A$ and $L$ have weak dependence on $T_e$,

$$\frac{V_{col,mod}}{V_{col,exp}} = \frac{R_{col,mod}}{R_{col,exp}} = \frac{\sigma_{col,exp}(n_{e,exp}, T_{e,exp})}{\sigma_{col,mod}(n_{e,mod}, T_{e,mod})} \equiv \delta(n_{e,exp}, T_{e,exp})$$

Because the theoretical discrepancy in the column voltage is $\Delta V_{col} = V_{col,exp} - V_{col,mod}$, the quantity $\Delta V_{col}$ can be expressed in terms of $\delta$,

$$\Delta V_{col}(n_e, T_e) = \left(\frac{1 - \delta(n_e, T_e)}{\delta(n_e, T_e)}\right) V_{col,mod}$$

The total discharge voltage can be written as $V_d = V_{a,sh} + V_{col} + V_{c,sh}$, where $V_{a,sh}$ and $V_{c,sh}$ are the anode and cathode sheath voltage drops, respectively, and $V_{col}$ is the column voltage drop.

The $I_d = 55$A, 1.5 mm gap case in [16] is compared with the $I_d = 55$ A, 2.0 mm gap case in this work. Specifically, the quantity $V_{d,mod} + \Delta V_{col}$ is calculated for this case and compared to the independently measured column drop in this work $V_{col,exp}^*$ (

Table 4). The theoretical discrepancy due to column conductivity decrease, $\Delta V_{col}$, is calculated assuming values for $T_e$ acquired using the pLTE OES method and $n_e$ from Stark broadening method. In this case, the experimental discharge voltage is found to be $18.4 \pm 0.3$ V while $V_{d,mod} = 10.5$ V, giving a discrepancy of 7.9 V (Figure 17). Apparently, $\Delta V_{col} = 1.7 \pm 0.8$ V, accounting for 20% of the discrepancy in discharge voltage between Ref. [16] and the discharge voltage observed in this work. The indicates that a portion of the discharge voltage underprediction may be accounted for by the model's overprediction of $T_e$ and $n_e$. The remaining $6.2 \pm 0.9$ V discrepancy may be accounted for by an underprediction in the anode and cathode sheaths found in Ref. [16] and this work.

| $V_{col,mod}$ (V) | $\Delta V_{col}$ (V) | $V_{col,exp}^*$ (V) | $V_{col,mod} + \Delta V_{col}$ (V) |
|---|---|---|---|
| 2.9 | $1.7 \pm 0.8$ | $4.6 \pm 0.3$ | $4.6 \pm 0.8$ |

**Table 4.** Arc column drops from Ref. *[16]* and as measured in this paper, as well as the theoretical increase in column drop due to $n_e$ and $T_e$ underprediction, $\Delta V_{col}$.

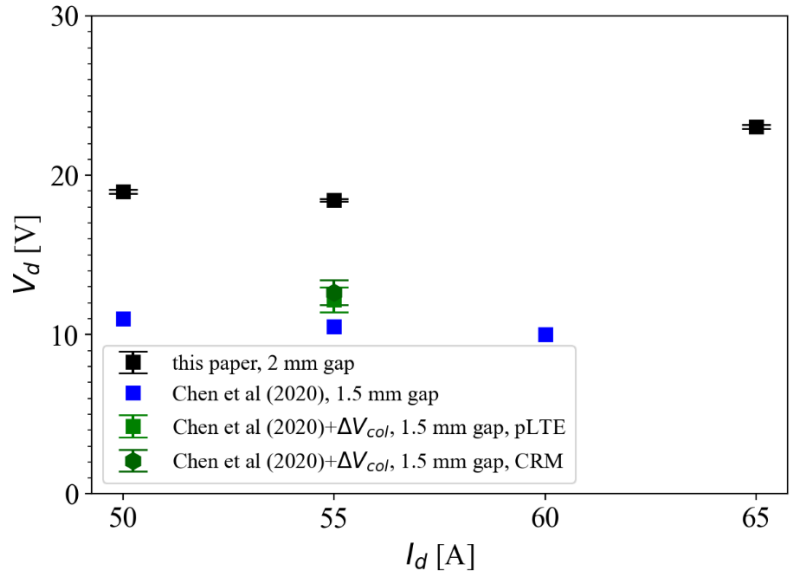

**Figure 17.** IV of carbon arc from this work and in Ref. *[16]* as well as the predicted voltage discrepancy between the model in Ref. *[16]* and this paper's results due to difference in modeled vs. measured $T_e$ and $n_e$.

## 6. Conclusions

The results of this paper are relevant to the design and implementation of future ablative nanosynthesis reactors. The anode sheath in an anodic carbon arc for synthesis of nanomaterials was investigated using electrostatic probe and spectroscopic techniques. The ablation of the anode serves as a feedstock of carbon for production of nanomaterials in the arc. The rate at which the anode ablates depends on the thermal flux of electrons to the anode surface, which can be modulated by the anode sheath. In this work, the anode sheath was determined to be positive in both low and high ablation regimes. Moreover, the positive anode sheath increases from low to high ablation rate regimes. The measured anode sheath and ablation rate in the high ablation mode is compared to a recently developed model of the anode ablation rate [15]. The ablation model agrees with the anode ablation rate if only a fraction of the anode sheath energy delivered to electrons is spent on heating the anode surface directly. This suggests that a fraction of the power delivered by the anode sheath may be lost to dissociating carbon material liberated from the anode surface.

    The electron temperature was found to be significantly smaller than in Ref. [14], likely due to the usage of a different collisional radiative model in determining $T_e$ via OES. In this work, calculations of $T_e$ were compared between the pLTE assumption and a CRM, both of which converge to the same values of the electron temperature within experimental error. The electron temperature and electron density determined in this work have uncovered a plausible cause for the discrepancy of discharge voltage in recent models with experimentally observed discharge voltage. Because the $T_e$ used as a fitting parameter in Ref. [16] was larger than experiment, the plasma column conductivity was lower and hence the modeled discharge voltage underpredicted the experimentally observed discharge voltage. The modeled column voltage matches experiment within error when using the $T_e$ and $n_e$ found in this work to determine plasma column conductivity.

# 7. Acknowledgements

This work was performed under the U.S. Department of Energy through contract DE-AC02-09CH11466. The authors are grateful to Jian Chen and Alexander Khrabry for insight into modeling of the carbon arc, Igor Kaganovich and Valerian Nemchinsky for fruitful discussions on anode sheath phenomena, and Jacob Simmonds for fruitful discussions and technical support on probe diagnostics.

## 9. Appendix

### A. Probe models

#### a. Presheath models

Recall the arc anode sheath $V_{a,sh}$ is determined as

$$V_{a,sh} = V_a - V_{pl,a}$$
$$V_{pl,a} = V_{pr} + V_{sh} + V_{pre} + V_{col}$$

In this work, different models of the presheath are considered, all of which can be written as a factor multiple of the electron temperature:

$$V_{pre} = \alpha \frac{kT_e}{e}$$

For the collisionless presheath, $\alpha = \frac{1}{2}$. By adding the effects of collisions into the presheath, it will be found that $\alpha > 1/2$, increasing the difference between the probe potential and the plasma potential. The equation for the anode sheath is then rewritten as

$$V_{a,sh} = V_a - V_{fl} - \left(\frac{1}{2} T_e \ln\left(\frac{M_i}{2\pi\, m_e}\right) + \alpha\right) T_e$$

For a fixed anode voltage and for a measured probe floating potential, an increased presheath contribution makes the anode sheath prediction smaller. Therefore, adding collisions will make a more conservative (more negative) calculation of the anode sheath drop. It will be shown that despite this, the anode sheath drop is positive in both low and high ablation regimes.

Following the model in Ref. [22] for a presheath within which ions perform many collisions with background neutrals, a spatial variation of the potential away near a wall can be determined by solving the transcendental equation

$$x = \frac{1}{2}(1 - e^{-2\chi} - 2\chi) \tag{12}$$
$$\chi = -\frac{eU(z)}{kT_e}$$
$$x = \frac{z}{L}$$
$$L = \lambda_{i,MFP,mom.}$$

where $z$ is the distance from the wall, $\lambda_{i,MFP,mom.}$ is the ion momentum transfer mean free path, and $U(z)$ is the potential at $z$ relative to the potential at the sheath-presheath boundary. Since a thin sheath is assumed, the sheath-presheath boundary is assumed to be at $z = 0$.

After many $L$ away from the probe, the presheath drop will be infinite. However, the maximum presheath drop can be estimated to be one interelectrode spacing away from the probe, since that is the largest length scale in the arc system. The presheath predicted by case (b) in Riemann (1991) increases in magnitude monotonically with distance from the probe wall. The maximum presheath drop would be 4.5 mm from the probe wall, because the probe measurements in this work are performed in a 4.5 mm interelectrode gap. Therefore, a maximum presheath drop calculated using $z = 4.5$ mm yields a minimum estimate of the positive anode sheath.

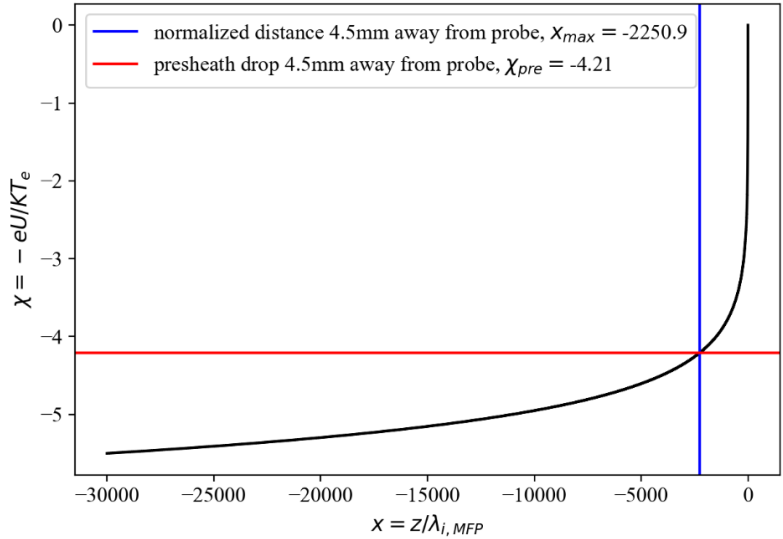

**Figure 18.** Plot of normalized presheath potential drop vs. normalized distance from probe wall, using $n_e$ and $T_e$ found from OES for a 4.5 mm gap, 50 A arc. The solution 4.5 mm from the probe wall is designated by the blue and red lines.

As an example, performing the calculation for $n_e$ and $T_e$ observed for the 50A arc, the presheath drop is determined to be $V_{pre} = U(z = 4.5 \text{ mm}) = 4.21 T_e$ (Figure 18). With a collisionless sheath, the full drop from a carbon plasma to probe wall is then $V_{fl} = V_{sh} + V_{pre} = 8.29 \frac{kT_e}{e}$.

### b. Nemchinsky (2009) model of ion saturation current compared to classical Bohm current

Ref. [27] shows that the ion current density to a probe can be expressed as in Eq. (7). Comparing the determination of $n_e$ from this model to that of the classical Bohm ion saturation current $I_{i,sat,Bohm} = 0.61 n_e \sqrt{kT_e/m_i} A_{pr}$, the model in Ref. [27] of ion saturation current provides a better agreement to the Stark broadening method performed with OES in this and previous works [14]. This suggests that the effects of ionization and ion collisions with neutrals in the probe presheath are important for probe diagnostics in the atmospheric pressure carbon arc.

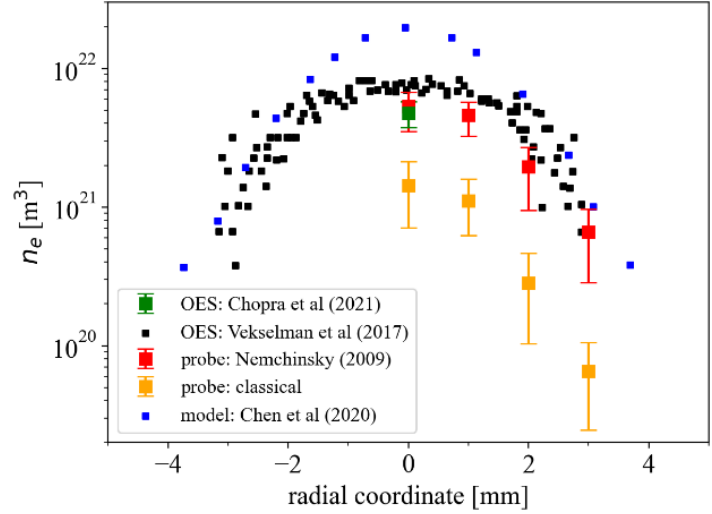

**Figure 19.** Plasma density as determined by OES and ion saturation current data, including different models of ion saturation current.

### B. Spectroscopic considerations

#### I. Plasma opacity

The optical depth of the spectroscopic line $\lambda_0$ is [29]

$$\tau(\lambda_0) = \pi r_e \lambda_0 f_{ik} n_i d \sqrt{\frac{Mc^2}{2\pi k_B T_a}} \tag{13}$$

where $r_e$ is the classical electron radius, $f_{ik}$ is the absorption oscillator strength, $M$ is the absorber (hydrogen) mass, $c$ is the speed of light, and $d = 2.0$ mm is the largest length of the absorbing slab of plasma (taken to be the arc radius) along the optical line of sight. The number density of atoms in the lower level of the transition (in the case of the Balmer series, hydrogen atoms in the $n = 2$ excited state) is given by a Boltzmann distribution $n_i = n_a \exp(-E_i/k_B T_a)$. Here $n_a$ is the number density of absorbers (hydrogen neutrals) calculated as a percentage in concentration of the total gas mixture (500 torr total background gas pressure, 95% He, 5% $H_2$ by concentration). The absorber temperature $T_a$ is assumed to be the same as the gas temperature. Calculating the opacity to the Balmer series lines, it is found that $\tau \ll 1$ in the range of $T_e$ applicable to the carbon arc, meaning the effects of reabsorption on emitted Balmer series photons is negligible.

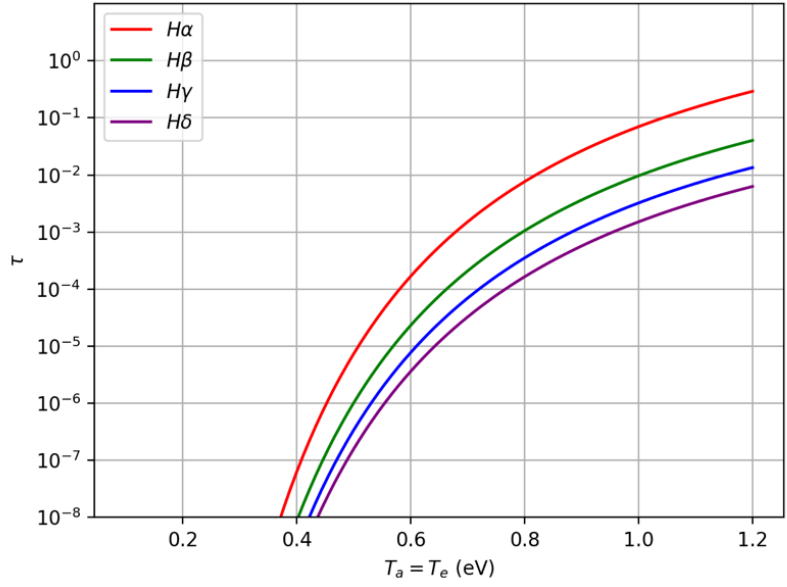

**Figure 20.** Plot of the optical depth $\tau$ vs. the electron temperature $T_e$ for the first four Balmer series lines.

## II. OES data analysis methods

*Boltzmann diagram method*

The first method to determine electron temperature was the Boltzmann diagram method assuming pLTE. A typical raw OES data acquisition and extracted line profile in the arc core (radial coordinate $r = 0$ mm) is shown in Figure 21. $T_e$ is determined by performing a linear fit to the corrected line intensities corresponding to transition from upper state $p$ to lower state $q$ as shown by the following equation:

$$\ln\left(\frac{I(p \to q)\lambda(p \to q)k_{BB}(p \to q)}{A(p \to q)g(p)}\right) = \frac{E(p \to 1)}{k_B T_e} + C \qquad (14)$$

Here $I(p \to q)$ is the measured intensity for the $p \to q$ transition, $\lambda(p \to q)$ is the corresponding wavelength, $k_{BB}(p \to q)$ is the blackbody correction factor, $A(p \to q)$ is the Einstein coefficient for spontaneous emission, $g(p)$ is the degeneracy of the $p$ state, $E(p \to 1)$ is the energy of the $p$ state referenced to the ground state, and $C$ is a constant offset.

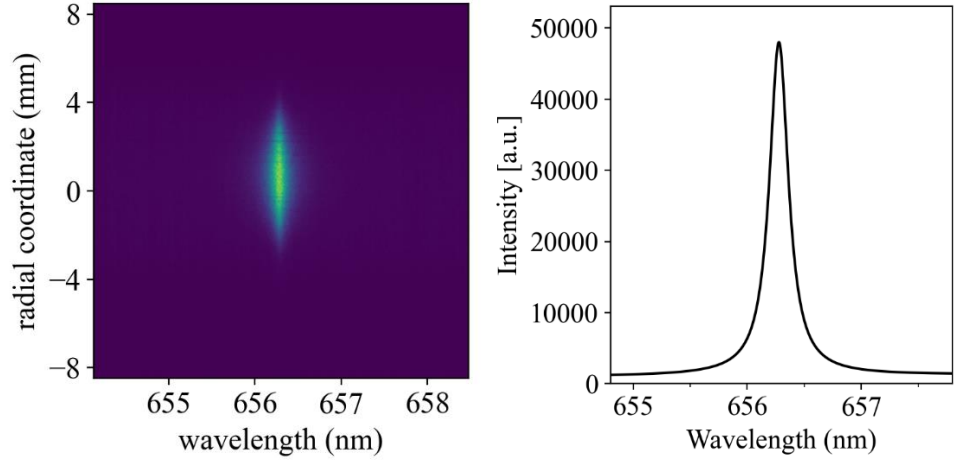

**Figure 21.** Typical acquisition of the $H_\alpha$ line via OES (50A, 4.5 mm gap) (left). Typical $H_\alpha$ line at radial coordinate $r = 0$ mm, (50A, 4.5 mm gap) (right).

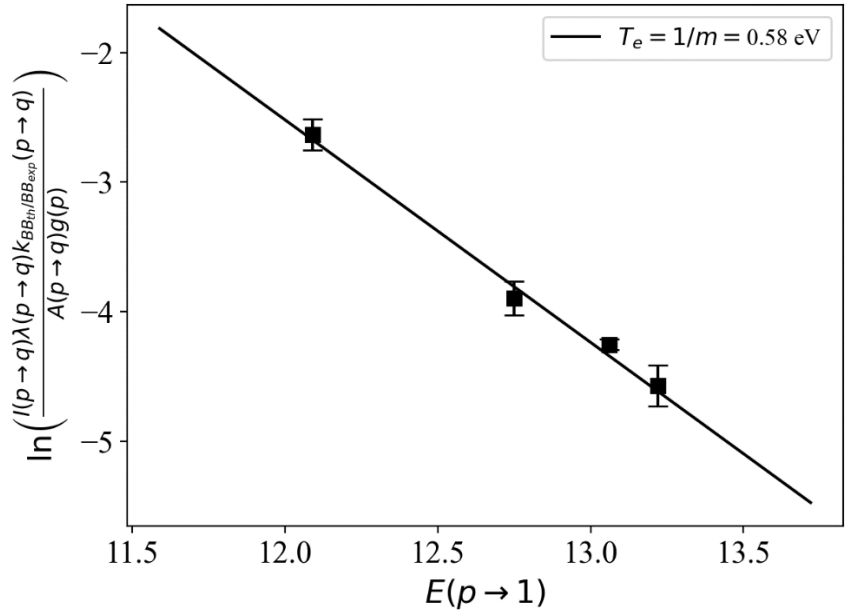

**Figure 22.** Example Boltzmann diagram (50A, 4.5 mm gap)

$T_e$ determined by Boltzmann diagram method was thus found to be roughly constant between low and high ablation regimes, $T_e = 0.58 \pm 0.06$ eV in the low ablation regime and $T_e = 0.65 \pm 0.07$ eV in the high ablation regime. $T_e$ was then used to determine the various broadening contributions to the $H_\alpha$ line. To determine the validity of the Boltzmann diagram method in determining $T_e$, the following criterion (Kunze) should be considered:

$$\frac{n_e}{m^{-3}} \geq 1.1 \times 10^{24} \frac{(z+1)^6}{n_{th}^{17/2}} \left(\frac{k_B T_e}{eV}\right)^{1/2} \tag{15}$$

The lowest energy level state considered in the analysis is the $n = 2$ state, therefore $n_{th} = 2$. Using these parameters, the $n \geq 2$ states are in pLTE if $n_e \geq 3 \times 10^{21}$ m$^{-3}$.

*Collisional radiative model*

To more accurately model the Balmer series emission, the role of general populating/de-population collisional mechanisms using a CRM is investigated. In this model, a set of $N$ total $nl$-atomic levels with radiative and collisional couplings is employed. These collisional interactions include:

- Radiative decay/absorption: $[A_{ml' \to nl} / A_{nl \to ml'}]$
- Electron-impact excitation/deexcitation: $[q^e_{ml' \to nl} / q^e_{nl \to ml'}]$
- Electron-impact ionization: $[S^e_{nl}]$
- Radiative recombination: $[\alpha^{(r)}_{nl}]$
- Three-body recombination: $[\alpha^{(3)}_{nl}]$

The rate equation for each of the excited populations of the $nl$-level of an $H$ atom is written in the form

$$\frac{dn_{nl}}{dt} = \sum_{ml' \neq nl} [A_{ml' \to nl} + n_e q^e_{ml' \to nl}] \cdot n_{ml'} \qquad (16)$$
$$- \left\{ n_e S^e_{nl} + \sum_{ml' \neq nl} [A_{nl \to ml'} + n_e q^e_{nl \to ml'}] \right\} \cdot n_{nl}$$
$$+ n^+_H \cdot n_e \cdot [\alpha^{(r)}_{nl} + n_e \alpha^{(3)}_{nl}]$$

Where $n_e$ and $n^+_H$ are the electron and hydrogen-ion densities, $A_{ml' \to nl}$ and $A_{ml' \to nl}$ are the Einstein coefficients for spontaneous emission and absorption, $S^e_{nl}$ is the electron-impact ionization volumetric rate coefficient, $q^e_{ml' \to nl}$ and $q^e_{nl \to ml'}$ are the electron-impact deexcitation and excitation volumetric rate coefficients, and $\alpha^{(r)}_{nl}$ and $\alpha^{(3)}_{nl}$ are the radiative and three-body recombination volumetric rate coefficients.

The couplings between the atomic states are succinctly written in terms of the collisional radiative matrix in the form

$$\frac{dn_{nl}}{dt} = \sum_{ml' \neq nl} (C_{nl,ml'} \cdot n_{ml'}) + C_{nl,nl} \cdot n_{nl} + n^+_H \cdot n_e \cdot R^e_{nl} \qquad (17)$$

Where the non-diagonal terms that include the populating matrix elements of the $nl$-th level are defined as

$$C_{nl,ml'} = A_{ml' \to nl} + n_e q^e_{ml' \to nl} \qquad (18)$$

and the diagonal (depopulating) elements as

$$C_{nl,nl} = -\left\{n_e S_{nl}^e + \sum_{ml' \neq nl} [A_{nl \to ml'} + n_e q_{nl \to ml'}^e]\right\} \quad (19)$$

The matrix diagonal is offset by the recombinative element

$$R_{nl}^e = \left[\alpha_{nl}^{(r)} + n_e \alpha_{nl}^{(3)}\right] \quad (20)$$

Assuming that the atomic relaxation times are small due to high electron densities found in the arc discharge, the quasi-static equilibrium assumption is made ($dn_{nl}/dt = 0$) and the solution is written in the form

$$n_{nl} = -n_H \cdot \sum_{ml'=1}^{N-1} \left(C_{nl,ml'}^{(r)}\right)^{-1} \cdot C_{ml'+1,nl}$$
$$-n_H^+ \cdot n_e \cdot \sum_{ml'=1}^{N-1} \left(C_{nl,ml'}^{(r)}\right)^{-1} \cdot R_{ml'+1}^e \quad (21)$$

Where $C_{nl,ml'}^{(r)}$ is the reduced collisional radiative matrix as described in [35] and $n_H$ is the total hydrogen neutral density. The rate equation for the neutral hydrogen density is given by

$$\frac{dn_H}{dt} = -n_e S^e \cdot n_H + n_e R^e \cdot n_H^+ \quad (22)$$

By normalizing the ionization balance using $n_H + n_H^+ = 1$, the solution to Eq. (22) is

$$n_H(t) = \frac{1}{R^e + S^e}\left(R^e + S^e e^{t/\tau_i}\right) \quad (23)$$

where the ionization balance relaxation time is given as $\tau_i = [n_e(R^e + S^e)]^{-1}$. Since the measured $n_e$ in the carbon arc is large $n_e > 10^{21}$ m$^{-3}$, the ionization balance relaxation time is small, justifying the assumption $\frac{dn_H}{dt} \approx 0$. Therefore, the normalized ionization balance solutions for neutrals and ions are given by

$$n_H = \frac{R^e}{R^e + S^e} \quad (24)$$

$$n_H^+ = \frac{S^e}{R^e + S^e}$$

These normalized solutions are substituted into Eq. (21) to obtain the quasi-static equilibrium atomic populations. The CRM normalized photo-emissivities in units of $[\text{s}^{-1}\text{sr}^{-1}]$ are then calculated using

$$\varepsilon_{nl \to ml'} = \frac{A_{nl \to ml'}}{4\pi} \cdot n_{nl} \qquad (25)$$

The emissivity ratios between different hydrogen lines calculated using the CRM are compared to the ratios calculated using the partial local thermodynamic equilibrium (pLTE) solution given by

$$\left(\varepsilon_{i \to j}\right)_{pLTE} \propto A_{i \to j} g_i e^{-E_i/kT_e} \qquad (26)$$

The atomic data employed in this model consists of state-of-the-art electron-impact excitation data using R-Matrix with Pseudo-States (RMPS) data [36,37] and convergent close-coupling (CCC) for electron-impact ionization [38]. The model includes up to the $n = 5$ shell of hydrogen and the high $n$-shell contributions are projected into the $n = 1 \to 5$ by means of the projection matrix using ADAS atomic subroutines [39].

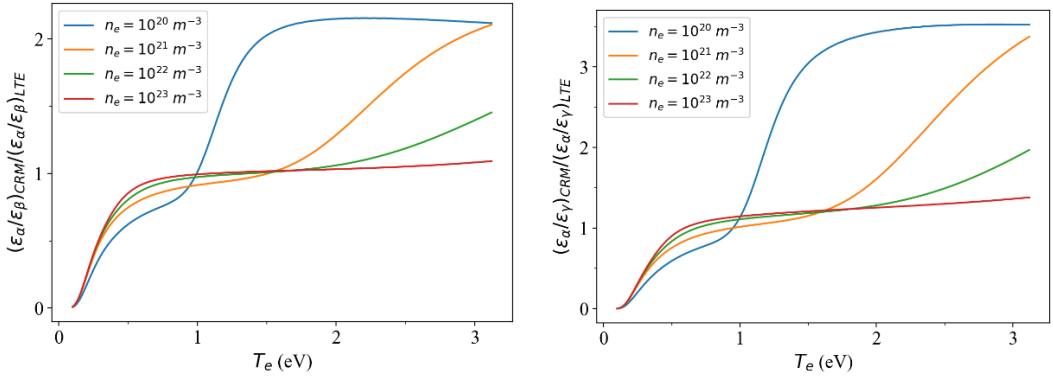

**Figure 23.** Comparisons between line-ratios calculated using the CRM to those from the LTE model for different temperatures and densities that are typical of the plasma discharge. Shown are ratios of CRM and pLTE emissivity ratios of $H_\alpha$ and $H_\beta$ lines (left) and $H_\alpha$ and $H_\gamma$ (right).

In order to compare results between emissivity ratios calculated using the CRM to those using the pLTE model, the emissivities for $H_\alpha$, $H_\beta$, and $H_\gamma$ lines are calculated using the CRM, and the ratios between the CRM calculated emissivities are compared to the emissivity ratios calculated from the pLTE model over a range of $T_e$ and $n_e$. As expected, for low $T_e$ the results rapidly approach the LTE values since $n_H^+ \approx 0$ due to the lack of ionization, and therefore the atomic populations approach LTE due to pure excitation. However, as $T_e$ increases, ionization increases and hence the contributions to the populations from both excitation and recombination enter the collisional-radiative regime. The LTE conditions are again obtained as $n_e$ increases.

*Determination of $n_e$ via Stark broadening*

The plasma density is determined by examining the spectral linewidth of the $H_\alpha$ line. The $H_\alpha$ line is broadened due to the above calculated factors: instrumental, Doppler, resonance, Van der Waals, and Stark broadening. Line broadening due to instrumental and Doppler broadening is Gaussian in nature, while resonance, Van der Waals, and Stark broadening are Lorentzian forms of broadening. Because the $H_\alpha$ line is broadened by a convolution of Gaussian and Lorentzian sources, the $H_\alpha$ FWHM line width is accordingly determined using a Voigt profile with FWHM of $\Delta\lambda_V$. Denote the instrumental broadening FWHM contribution as $\Delta\lambda^{instr.}$ and the Doppler broadening FWHM as $\Delta\lambda^D$. Similary, denote the Stark broadening FWHM as $\Delta\lambda^S$, Van der Waals broadening FWHM as $\Delta\lambda^{VdW}$, and resonance broadening FWHM as $\Delta\lambda^{res.}$. Gaussian broadening adds in quadrature and the Lorentzian broadening adds linearly, hence the total Lorentzian and Gaussian FWHM contributions can be written as

$$\Delta\lambda^L = \Delta\lambda^S + \Delta\lambda^{VdW} + \Delta\lambda^{res.}$$

and

$$\Delta\lambda^G = \sqrt{(\Delta\lambda^{instr.})^2 + (\Delta\lambda^D)^2}$$

Given $\Delta\lambda^L$ and $\Delta\lambda^G$, an approximation to $\Delta\lambda^V$ accurate to 0.02% is given by:

$$\Delta\lambda^V \approx 0.5346\,\Delta\lambda^L + \sqrt{0.2166\,(\Delta\lambda^L)^2 + (\Delta\lambda^G)^2}$$

After determining $\Delta\lambda^G$, the transcendental equation for $\Delta\lambda^L$ can be solved for in terms of $\Delta\lambda^G$ and $\Delta\lambda^V$. The solution, $\Delta\lambda^L = (\Delta\lambda^L)_{sol.}$ gives

$$\Delta\lambda^S = (\Delta\lambda^L)_{sol.} - \Delta\lambda^{VdW}$$

Finally, using the diagnostic maps calculated in [40] using a value of $\mu = 0.90$ corresponding to hydrogen emitters and carbon perturbers, for a determined $T_e$ and $\Delta\lambda^S$ a value can be extracted for $n_e$.

The instrumental broadening $\Delta\lambda^{instr.}$ is determined by directly imaging an Hg(Ar) spectral calibration lamp with the spectrometer and iCCD camera. The FWHM of the 579.1 nm Hg spectral line was used as the value of the instrumental broadening FWHM, which was found to be $\Delta\lambda^{instr.} = 25$ pm.

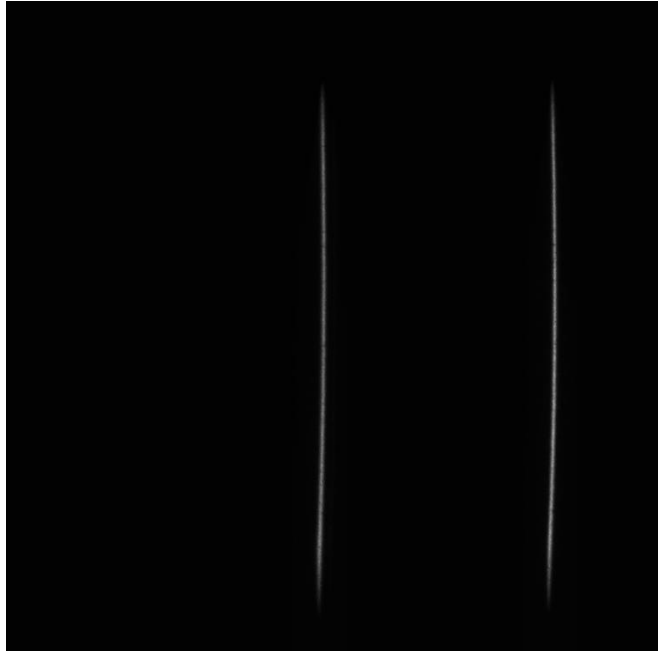

**Figure 24.** Spectral image of the 579.1 nm Hg line captured from the Hg(Ar) spectral calibration lamp.

Assuming the plasma is in thermal equilibrium, implying $T_H = T_e$ where $T_H$ is the temperature of the hydrogen atoms in the arc, the Doppler broadening FWHM is given by [41]

$$\Delta\lambda^D = \lambda_{3\to 2}\sqrt{8\ln 2\frac{k_B T_H}{m_H c^2}}$$

where $\lambda_{3\to 2}$ is the wavelength of the photon emitted by the $n = 3$ to 2 hydrogen atomic transition, $m_H$ is the hydrogen atomic mass, and $c$ is the speed of light. Calculated as such, the Doppler broadening contribution is found to be $\Delta\lambda^D = 36 - 44$ pm.

Van der Waals broadening is calculated using the expression given in [41], assuming the emitters are hydrogen atoms and the perturbing atoms are carbon atoms

$$\frac{\Delta\lambda^{VdW}}{\lambda_{3\to 2}} \approx 8.5 \times 10^{-17}\frac{\lambda_{3\to 2}}{\text{nm}}\left(\frac{C_6^{3\to 2}}{\text{m}^6\text{s}^{-1}}\right)^{\frac{2}{5}}\left(\frac{T/K}{\mu/u}\right)^{\frac{3}{10}}\left(\frac{n_C}{\text{m}^{-3}}\right)$$

$$C_6^{3\to 2} = \frac{e^2}{4\pi\varepsilon_0}\frac{1}{\hbar}\alpha_d|\langle R_3^2\rangle - \langle R_2^2\rangle|$$

Here $n_C$ is the carbon neutral density, taken to be $n_C = (500\text{ torr})/(k_B T)$, $e$ is the electron charge, $\varepsilon_0$ is the dielectric constant, $\hbar$ is the reduced Planck constant, and the interaction constant $C_6^{3\to 2}$ is calculated using the dipole polarizability $\alpha_d$ of a carbon atom described in [41,42]. The atomic mean squared radii of the $n = 3$ and 2 states, $\langle R_3^2\rangle$, $\langle R_2^2\rangle$, are approximated by the expression given in [5]. Found this way, $C_6 \approx 10^{-42}$ m$^6$s$^{-1}$, and hence $\Delta\lambda^{VdW} = 31 - 40$ pm.

The expression for resonance broadening given in [41] is used to calculated $\Delta\lambda^{res.}$

$$\Delta\lambda^{res} = 9 \times 10^{-34}\sqrt{\frac{g(g)}{g(p)}}f(g\to p)\frac{\lambda_{pq}}{\text{nm}}\frac{\lambda_{pg}}{\text{nm}}\frac{n_a(g)}{\text{m}^{-3}}\lambda_{pq}$$

where again $p = 3$, $q = 2$, $g = 1$ refers to the ground state of hydrogen, $f(g \to p)$ is the corresponding oscillator strength, the degeneracy of state $k$ is denoted as $g(k)$, and the density of ground state atoms associated with the emitters $n_a(g)$. Here, $n_a(g)$ represents ground state hydrogen atom emitters, the "worst case scenario" of maximum emitters is taken, where all hydrogen atoms are assumed to be in the ground state. Calculated as such, the resonance broadening is found to be negligible, with $\Delta\lambda^{res.} = 0.2 - 0.4$ pm.

| $\Delta\lambda^D$ (nm) | $\Delta\lambda^{res.}$ (nm) | $\Delta\lambda^{VdW}$ (nm) | $\Delta\lambda^{instr.}$ (nm) |
|---|---|---|---|
| 0.036-0.044 | 0.00026-0.00039 | 0.031-0.040 | 0.025 |

Table 5. Summary of FWHM contributions to H$_\alpha$ line broadening

## C. Determination of voltage drop over plasma column

In Shashurin et al (2008) investigation of the IV characteristic of a carbon arc with a hollow carbon anode composed of C:Ni:Y = 56:4:1, the discharge voltage was found to indeed increase with increasing gap size. This increase is attributed to the increase in the length of the interelectrode plasma column that has a relatively spatially homogenous electric field. Therefore, by obtaining discharge voltage measurements at two different interelectrode gap sizes at a fixed current, the electric field in the plasma column can be estimated. The plasma column electric field amplitude $E$ at a given discharge current $I_d$ can be estimated from the discharge currents at two different gap sizes $d_1$ and $d_2$ as

$$E(I_d) = \left| \frac{V_d(d_2) - V_d(d_1)}{d_2 - d_1} \right|$$

Assuming the probe tip measures the arc with an interelectrode gap $d$ at the arc midplane $d/2$, the drop over the column to the probe tip $V_{col}$ is then given as

$$V_{col} = E(I_d) \times \frac{d}{2}$$

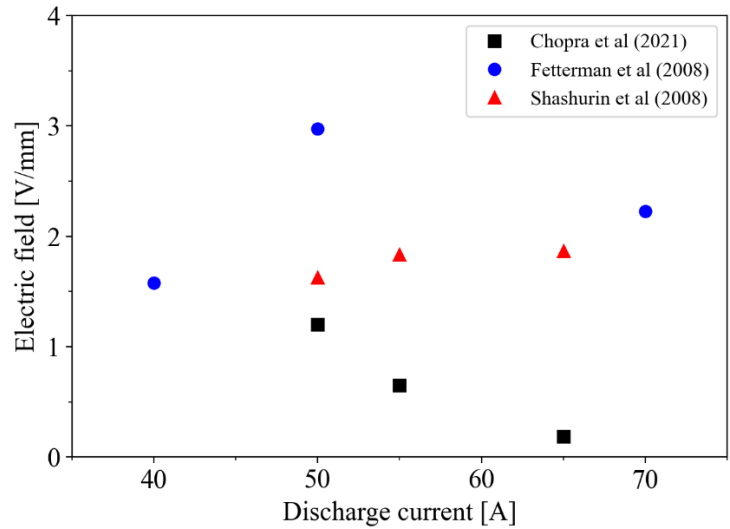

**Figure 25.** Electric field in the arc column calculated from IV data from different works.